\documentclass[a4paper,12pt]{article}
\usepackage{amsmath}
\usepackage{amsfonts}
\usepackage{amssymb}
\usepackage{bbm}
\usepackage{mathrsfs}
\usepackage{ulem}
\usepackage{epsfig}
\usepackage{wrapfig}
\usepackage{color}
\usepackage{type1cm}
\usepackage{cite}
\usepackage{hyperref}
\providecommand{\U}[1]{\protect\rule{.1in}{.1in}}
\numberwithin{equation}{section}

\newcommand{\rT}{\mathrm{T}}
\newcommand{\mGamma}{{\mit\Gamma}}
\newcommand{\doi}[1]{%
  \href{https://doi.org/#1}{doi:#1}%
}

\allowdisplaybreaks[2]

\begin{document}

\title{\textsf{5D Runge-Lenz Vector
    from Symplectic Reduction of 8D Harmonic Oscillator}}

\author{\textsf{Hitoshi Ikemori } \thanks{ikemori@vega.aichi-u.ac.jp}\\Faculty of Economics, Aichi University, \\Nagoya, Aichi 453-8777, Japan
\and \textsf{Shinsaku Kitakado }\thanks{kitakado@meijo-u.ac.jp}\\Professor Emeritus, Nagoya University
\and \textsf{Yoshimitsu Matsui }\thanks{matsui@vega.aichi-u.ac.jp}\\Faculty of Law, Aichi University,\\Nagoya, Aichi 453-8777, Japan
\and \textsf{Toshiyuki Okubo }\thanks{tokubo@meijo-u.ac.jp}\\Faculty of Science and Technology, Meijo University,\\Nagoya, Aichi 468-8502, Japan
\and \textsf{Toshiro Sato }\thanks{sttshr.aooio@gmail.com} \,\textsuperscript{,}\thanks{Now retired from the university.} \\School of Engineering, Chukyo University,\\Nagoya, Aichi, 466-8666, Japan}
\date{}
\maketitle

\begin{abstract}
  We investigate the origin of the Runge-Lenz vector
  of the five-dimensional Kepler problem in the SU(8) dynamical symmetry
  of the eight-dimensional harmonic oscillator acting on its phase space.
  Using a complete basis of real $8\times 8$ matrices
  constructed from five-dimensional gamma matrices
  and the $\mathfrak{su}(2)$ generators associated with
  the symplectic reduction, we decompose the conserved quantities
  of the harmonic oscillator into representations of
  $\mathrm{SO}(5) \times \mathrm{SU}(2)$ and identify the SU(2)-singlet sector.
  We show that the corresponding conserved quantities generate
  an SO(6) subgroup of the SU(8) dynamical symmetry.
  We then perform the Hurwitz transformation and symplectic reduction
  with respect to SU(2), obtaining the five-dimensional Kepler Hamiltonian.
  We explicitly show that the SO(6) moment map reduces to
  the angular momentum and the suitably normalized Runge-Lenz vector
  of the five-dimensional Kepler problem on a fixed negative-energy surface.
  This establishes the origin of the Runge-Lenz vector
  in the dynamical symmetry of the eight-dimensional harmonic oscillator.
\end{abstract}
  
\section{Introduction}

The Kepler problem~\cite{CordaniBook}, characterized by a $1/r$ potential,
is of fundamental importance in both classical and quantum mechanics.
In addition to the angular momentum associated with spatial rotations,
it possesses a conserved quantity
known as the Runge-Lenz (RL) vector~\cite{Runge,Lenz}.
In classical mechanics, the RL vector is closely related to
the fact that bound orbits are closed ellipses.
In quantum mechanics, it plays a central role in explaining
the degeneracy of the hydrogen-like atom that
cannot be accounted for by rotational symmetry alone.
Thus, the RL vector characterizes the dynamical symmetry
intrinsic to the Kepler system.
For negative energies, its suitably normalized form combines with
the angular momentum to enlarge the SO(3) rotational symmetry
to SO(4) dynamical symmetry.

In quantum mechanics,
one geometrical interpretation of this SO(4) dynamical symmetry
was provided by Fock and Bargmann
based on the geometry of momentum space~\cite{Fock,Bargman}.
In their approach, for a fixed negative energy,
the momentum space is mapped onto a three-dimensional sphere,
thereby elucidating the geometrical origin of its SO(4) dynamical symmetry
and the role of the RL vector.
A closely related geometrical picture appears
in the classical Kepler problem.
Moser regularized the Kepler flow at fixed negative energy
by relating it, through stereographic projection,
to the geodesic flow on a three-dimensional sphere~\cite{Moser}.
This construction was further developed by Ligon and Schaaf,
who gave a canonical regularization
of the negative-energy Kepler phase space
in which the SO(4) dynamical symmetry is globally realized~\cite{LS}.

Another interesting topic related to the Kepler problem is
its correspondence with the isotropic harmonic oscillator.
The relation between the two-dimensional harmonic oscillator
and the two-dimensional Kepler problem is established by
the Levi-Civita transformation~\cite{Levi-Civita},
while the four-dimensional harmonic oscillator
is mapped to the three-dimensional Kepler problem
by the Kustaanheimo-Stiefel transformation~\cite{KS}.
These transformations map harmonic-oscillator motion to Kepler motion
through a change of variables and a reparametrization of time.
The relation between the Kustaanheimo-Stiefel and Moser regularizations
of the three-dimensional Kepler problem has been investigated
from the viewpoint of symplectic reduction~\cite{Kummer}.
More recently, the three-dimensional Kepler problem has been formulated
as a geometric reduction of the four-dimensional harmonic oscillator,
with its conserved quantities related to the symmetries
of the harmonic oscillator~\cite{vanderMeer}.

In our previous work~\cite{IKMS}, we have seen,
using the formalism of geometric algebra, also known as Clifford algebra,
that the RL vector in the three-dimensional Kepler system
is related to the moment map of the SO(4) group on the phase space
of the four-dimensional particle motion.
This result motivates us to investigate
an analogous classical moment-map interpretation in higher dimensions,
since the RL vector is not restricted to the three-dimensional Kepler problem
but exists in arbitrary spatial dimensions~\cite{SMO}.

In higher dimensions, a transformation
from the eight-dimensional harmonic oscillator
to the five-dimensional Kepler problem,
known as the Hurwitz transformation,
has also been proposed and studied
by several authors~\cite{DMOST,LK,Cordani,LeK}.
The five-dimensional Kepler problem has also been investigated
from the viewpoint of symplectic reduction and dynamical symmetry.
The SU(2) Kepler problem has been constructed through
the symplectic reduction of the eight-dimensional conformal Kepler problem,
with the ordinary five-dimensional Kepler problem recovered
when the SU(2) moment map vanishes~\cite{Iwai}.
In the quantum formulation, its SU(4) symmetry for negative energies
was obtained by identifying the subgroup of the SU(8) symmetry
of the eight-dimensional harmonic oscillator that commutes with SU(2).
The corresponding generators were explicitly reduced to angular momentum
and Runge-Lenz-like operators~\cite{IS1}.
Furthermore, the dynamical group $\mathrm{SO}^\ast(8)$
and its associated moment map have been investigated
in a classical Hamiltonian framework,
leading to the construction of the conserved quantities
of the SU(2) Kepler problem~\cite{IS2}.

The correspondence between eight-dimensional harmonic
and singular oscillators and five-dimensional Kepler-type systems
with SU(2) monopoles has been further investigated,
including the construction of generalized Runge-Lenz vectors
and the analysis of SO(6) and SO(6, 2) symmetries~\cite{MST1,PT1,MST2,PT2}.
The reduction of eight-dimensional (pseudo)spherical harmonic oscillators
to five-dimensional Yang-Coulomb systems has also been studied~\cite{BTY}.

These studies have clarified various aspects
of the oscillator-Kepler correspondence
and the conserved quantities and dynamical symmetries of the associated systems.
In the present work, we investigate the origin of the RL vector
of the five-dimensional Kepler problem in the SU(8) dynamical symmetry
of the eight-dimensional harmonic oscillator acting
on its phase space $\mathbb{R}^8 \times \mathbb{R}^8$.
We begin by formulating the conserved quantities of the harmonic oscillator
as moment maps. Using a complete basis of real $8 \times 8$ matrices
constructed from five-dimensional gamma matrices
and the generators of $\mathfrak{su}(2)$ associated with
the symplectic reduction, we decompose these conserved quantities
into representations of $\mathrm{SO}(5) \times \mathrm{SU}(2)$
and identify the SU(2)-singlet sector.
The corresponding conserved quantities generate
an SO(6) subgroup of the SU(8) dynamical symmetry.
We then derive the five-dimensional Kepler problem
through the Hurwitz transformation and symplectic reduction
with respect to SU(2),
based on the Marsden-Weinstein (MW) theorem~\cite{MW}.
Through this reduction, we explicitly show that the SO(6) moment map
gives rise to the angular momentum and the suitably normalized RL vector
of the five-dimensional Kepler problem on a fixed negative-energy surface.

This paper is organized as follows.
In sections \ref{8DHO} and \ref{5DKepler}, we introduce moment maps
and describe the conserved quantities and dynamical symmetries of
the eight-dimensional harmonic oscillator
and the five-dimensional Kepler problem.
In section \ref{8x8completebasis}, we use five-dimensional gamma matrices
to extract from the conserved quantities
of the eight-dimensional harmonic oscillator
those that give rise to the conserved quantities
of the five-dimensional Kepler problem.
In sections \ref{derivation} and \ref{RL},
we derive the five-dimensional Kepler problem
by the Hurwitz transformation and symplectic reduction
and explicitly establish the origin of its RL vector.
In section \ref{summary}, we summarize our results.
The Fierz identities used in the derivations are collected
in appendix \ref{FierzIdentity}.
The interchange between energy and coupling constant 
in the transformation between the harmonic oscillator and the Kepler problem
is detailed in appendix \ref{interchange}.

\section{Conserved quantities of 8D harmonic oscillator}
\label{8DHO}

In this section, we review the conserved quantities of
the eight-dimensional harmonic oscillator
in terms of their corresponding moment maps.
The Hamiltonian on phase space
$T^\ast \mathbb{R}^8 \simeq \mathbb{R}^8 \times \mathbb{R}^8$,
with canonical coordinate $(q_a,\, p_a)$
and symplectic form $\omega = dq_a \wedge dp_a$, is given by
\begin{equation}
  h = \frac{1}{2} p^\rT p + \frac{1}{2} \lambda^2 q^\rT q , \label{8dimHOH}
\end{equation}
where $\lambda > 0$.

We first review the definition of the moment map~\cite{MR,Woit}.
Let $(M,\, \omega)$ be a symplectic manifold, and let $G$ be a Lie group
acting on $M$. A map
\begin{equation}
  \mu : M \to \mathfrak{g}^\ast
\end{equation}
is called a moment map if it satisfies the following conditions:
\begin{enumerate}
  \item $\mu$ is $G$-equivariant, namely,
  \begin{equation}
    \mu (g \cdot x) = \mathrm{Ad}_{g^{-1}}^\ast \mu (x), \quad
    \forall x \in M,\, g\in G .
  \end{equation}
  \item Let $X$ denote the tangent vector field generated by the action of $G$,
  \begin{equation}
    \iota_X \omega = d \mu .
  \end{equation}
\end{enumerate}
By Cartan's formula,
\begin{equation}
  {\cal L}_X \omega = d (\iota _X \omega) + \iota_X d \omega.
\end{equation}
Since $d\omega=0$, if $M$ is simply connected,
the second condition is equivalent to ${\cal L}_X \omega = 0$.
Thus, the symplectic form is invariant under the flow generated by $X$,
and the corresponding moment map serves as the generator
of the canonical transformation.

We now determine the moment maps associated with
the symmetries of the eight-dimensional harmonic oscillator.
Let $E^{(ab)}$ denote the standard matrix unit with a $1$ in the
$(a,\, b)$-entry and zeros elsewhere. For arbitrary vectors
$u_a,\, v_a \in \mathbb{R}^8$, it satisfies
\begin{equation}
  u^\rT E^{(ab)} v = u_a v_b .
\end{equation}
We then define the antisymmetric and symmetric matrices by
\begin{equation}
  A^{(ab)} = E^{(ab)} - E^{(ba)}, \quad S^{(ab)} = E^{(ab)} + E^{(ba)},
\end{equation}
respectively. The matrices $A^{(ab)}$ form a basis
of the $\mathfrak{so}(8)$ algebra.

We consider the SO(8) action generated by $A^{(ab)}$
on the phase space coordinates
$(q_a,\, p_a) \in \mathbb{R}^8 \times \mathbb{R}^8$ given by
\begin{equation}
  \left(
  \begin{array}{c}
    q \\ p
  \end{array} \right)
  \longmapsto
  \exp \left(
    \begin{array}{cc}
      - t A^{(ab)} & 0 \\
      0 & - t A^{(ab)}
    \end{array} \right) \left(
  \begin{array}{c}
    q \\ p
  \end{array} \right) .
\end{equation}
This action induces the tangent vector field
\begin{align}
  X &= \frac{d}{dt} = \left. \frac{dq_a}{dt} \right|_{t = 0}
    \frac{\partial}{\partial q_a}
    + \left. \frac{dp_a}{dt} \right|_{t = 0}
    \frac{\partial}{\partial p_a} \notag \\
  &= - q^\rT (A^{(ab)})^\rT \frac{\partial}{\partial q}
    - p^\rT (A^{(ab)})^\rT \frac{\partial}{\partial p} \notag \\
  &= q^\rT A^{(ab)} \frac{\partial}{\partial q}
    + p^\rT A^{(ab)} \frac{\partial}{\partial p} .
\end{align}
Contracting $X$ with the symplectic form $\omega$, we obtain
\begin{align}
  \iota_X \omega &= \omega (X,\, \cdot\,) \notag \\
  &= dq_a \wedge dp_a \left( q^\rT A^{(ab)} \frac{\partial}{\partial q}
    + p^\rT A^{(ab)} \frac{\partial}{\partial p} ,\, \cdot \, \right) \notag \\
  &= q^\rT A^{(ab)} dp - p^\rT A^{(ab)} dq \notag \\
  &= q_a dp_b - q_b dp_a - p_a dq_b + p_b dq_a \notag \\
  &= d( q_a p_b - q_b p_a) .
\end{align}
Hence, the corresponding moment map is given by
\begin{equation}
  \mu_{ab}^\mathrm{A} = q_a p_b - q_b p_a .
\end{equation}
This coincides with the angular momentum.
Thus, the moment map associated with this SO(8) action on the phase space
is the angular momentum.

Besides the SO(8) symmetry, the eight-dimensional harmonic oscillator
possesses a dynamical symmetry whose generators are the matrices $S^{(ab)}$.
We consider the following transformation generated by $S^{(ab)}$.
\begin{equation}
  \left(
  \begin{array}{c}
    \lambda q \\ p
  \end{array} \right)
  \longmapsto
  \exp \left(
    \begin{array}{cc}
      0 & t S^{(ab)}/2 \\
      - t S^{(ab)}/2 & 0
    \end{array} \right) \left(
  \begin{array}{c}
    \lambda q \\ p
  \end{array} \right) . \label{expS}
\end{equation}
The corresponding moment map is found to be
\begin{equation}
  \mu_{ab}^\mathrm{S} = \frac{1}{2 \lambda} \left( \lambda^2 q_a q_b
    + p_a p_b \right) .
\end{equation}
Decomposing this moment map into its trace and traceless parts as
\begin{align}
  \mu_{ab}^\mathrm{S} &= \left( \mu_{ab}^\mathrm{S}
    - \frac{1}{8} \delta_{ab} \mu_{cc}^\mathrm{S}\right)
    + \frac{1}{8}  \delta_{ab} \mu_{cc}^\mathrm{S} \notag \\
  &= \tilde{\mu}_{ab}^\mathrm{S} + \frac{1}{8} \delta_{ab} \mu_{cc}^\mathrm{S} .
\end{align}
Multiplying both the trace and traceless parts by $\lambda$,
we find that the trace part $\lambda \mu_{cc}^\mathrm{S}$ is the Hamiltonian,
while the traceless part $\lambda \tilde{\mu}_{ab}^\mathrm{S}$ provides
35 independent conserved quantities.
The symmetric tensor $\lambda \mu_{ab}^\mathrm{S}$
is proportional to the Jauch-Hill-Fradkin (JHF) tensor~\cite{JH,Fradkin}.
Together with the 28 generators of the SO(8) rotational symmetry,
the 35 traceless symmetric generators form the $\mathfrak{su}(8)$ algebra.
Thus, the eight-dimensional harmonic oscillator possesses
an SU(8) dynamical symmetry acting on its phase space $
\mathbb{R}^8 \times \mathbb{R}^8$.
Including the U(1) symmetry generated by the Hamiltonian,
the full group of linear canonical transformations preserving
the Hamiltonian is U(8).

\section{5D Kepler problem and RL vector}
\label{5DKepler}

We now turn to the five-dimensional Kepler problem and its conserved quantities.
The Hamiltonian on phase space with coordinate
$(x_i,\, \pi_i) \in \mathbb{R}^5 \times \mathbb{R}^5$ is given by
\begin{equation}
  H = \frac{1}{2} \pi_i^2 - \frac{k}{\sqrt{x_i^2}} , \label{5dimKeplerH}
\end{equation}
where $k > 0$.\footnote{%
More precisely, the origin of the configuration space
is excluded because the Kepler potential is singular there.}

The moment map associated with the SO(5) rotational symmetry
is given by the angular momentum
\begin{equation}
  L_{ij} = x_i \pi_j - x_j \pi_i .
\end{equation}

Another conserved quantity of the Kepler problem is the RL vector,
given by
\begin{align}
  A_i &= \pi_j L_{ij} - \frac{k}{\sqrt{x_k^2}} x_i \notag \\
  &= x_i \pi_k^2 - \pi_i (\pi_k x_k) - \frac{k}{\sqrt{x_k^2}} x_i .
\end{align}
Using Hamilton's equations and the conservation of the angular momentum,
we obtain
\begin{align}
  \frac{d}{dt} A_i &= L_{ij} \frac{d \pi_j}{dt}
    - k \left( \frac{1}{\sqrt{x_i^2}} \frac{d x_i}{dt}
    - \frac{x_i x_k}{(x_i^2)^{3/2}} \frac{d x_k}{dt} \right) \notag \\
  &= (x_i \pi_j - x_j \pi_i) \left( -k \frac{x_j}{(x_k^2)^{3/2}} \right)
    - k \left( \frac{\pi_i}{\sqrt{x_i^2}}
    - \frac{x_i x_k \pi_k}{(x_k)^{3/2}} \right) \notag \\
  &= 0 .
\end{align}
Unlike the angular momentum, the RL vector explicitly depends on
the coupling constant $k$ of the Kepler potential.
Its relation to a moment map is therefore less immediate.
However, in the region of phase space where $H < 0$,
we define the normalized RL vector $\tilde{A}_i$ by
\begin{equation}
  \tilde{A}_i = \frac{A_i}{\sqrt{- 2 H}} ,
\end{equation}
and identify $L_{i6} = \tilde{A}_i$.
Then, together with the SO(5) moment map $L_{ij}$,
the quantities $L_{IJ}$ $(I,\, J = 1,\ldots,6)$
satisfy the $\mathfrak{so}(6)$ algebra,
\begin{equation}
  \{ L_{IJ},\, L_{KL} \} = \delta_{IK} L_{JL} - \delta_{IL} L_{JK}
    - \delta_{JK} L_{IL} + \delta_{JL} L_{IK} . \label{so(6)algebra}
\end{equation}
This suggests that the normalized RL vector can be interpreted
as a component of an SO(6) moment map.
In the following section, we identify the corresponding conserved quantities
of the eight-dimensional harmonic oscillator.

\section{A complete basis for the space of $8\times 8$ matrices}
\label{8x8completebasis}

To identify the conserved quantities that survive the symplectic reduction,
we express the $8 \times 8$ antisymmetric and traceless symmetric matrices
associated with the eight-dimensional harmonic oscillator
in terms of five-dimensional gamma matrices and $\mathfrak{su}(2)$ generators.

First, we define a real representation
of the five-dimensional gamma matrices $\mGamma_i$ ($i = 1,\ldots,5$).
The explicit form is given, for instance, by
\begin{align}
  \mGamma_1 &= (i \sigma_2) \otimes \sigma_1 \otimes (i \sigma_2),
    \label{gamma1} \\
  \mGamma_2 &= - (i \sigma_2) \otimes (i \sigma_2) \otimes
    \mathbbm{1}_2, \label{gamma2} \\
  \mGamma_3 &= (i \sigma_2) \otimes \sigma_3 \otimes (i \sigma_2),
    \label{gamma3} \\
  \mGamma_4 &= \sigma_1 \otimes \mathbbm{1}_2
    \otimes \mathbbm{1}_2, \label{gamma4} \\
  \mGamma_5 &= \sigma_3 \otimes \mathbbm{1}_2
    \otimes \mathbbm{1}_2. \label{gamma5}
\end{align}
Here, $\sigma_i$ denote the Pauli matrices
\begin{equation}
  \sigma_1 = \left(
    \begin{array}{cc}
      0 & 1 \\ 1 & 0
    \end{array} \right) , \quad
  \sigma_2 = \left(
    \begin{array}{cc}
      0 & - i \\ i & 0
    \end{array} \right) , \quad
  \sigma_3 = \left(
    \begin{array}{cc}
      1 & 0 \\ 0 & -1
    \end{array} \right) .
\end{equation}
All the matrices given in (\ref{gamma1})-(\ref{gamma5})
are traceless symmetric matrices.
They satisfy the five-dimensional Clifford algebra
\begin{equation}
  \mGamma_i \mGamma_j + \mGamma_j \mGamma_i
    = 2 \delta_{ij} \mathbbm{1}_8 . \label{Clifford}
\end{equation}
Furthermore, the five gamma matrices satisfy
\begin{equation}
  \mGamma_1 \mGamma_2 \mGamma_3 \mGamma_4 \mGamma_5 = \mathbbm{1}_8 .
\end{equation}
Consequently, the antisymmetrized products of rank three or higher
can be expressed in terms of the identity matrix,
$\mGamma_i$, and $\mGamma_{ij} = \frac{1}{2} [\mGamma_i,\, \mGamma_j]$,
and therefore provide no additional independent matrices.

Next, we define the matrices $T_A$ ($A = 1,\,2,\,3$) by
\begin{align}
  T_1 &= - \mathbbm{1}_2 \otimes \mathbbm{1}_2 \otimes (i\sigma_2) ,
    \label{T1} \\
  T_2 &= - \mathbbm{1}_2 \otimes(i \sigma_2) \otimes \sigma_3 , \label{T2} \\
  T_3 &= \mathbbm{1}_2 \otimes (i \sigma_2) \otimes \sigma_1 . \label{T3}
\end{align}
They are antisymmetric and satisfy the relation
\begin{equation}
  T_A T_B = - \delta_{AB} \mathbbm{1}_8 - \varepsilon_{ABC} T_C ,
    \label{su2algebra}
\end{equation}
thereby forming the $\mathfrak{su}(2)$ algebra.\footnote{%
The explicit representations of $\mGamma_i$ and $T_A$ are equivalent,
up to relabeling, sign changes, and normalization,
to those used in ref.~\cite{BTY}.}
Moreover,
$\mGamma_i$ and $T_A$ commute
\begin{equation}
  [ \mGamma_i,\, T_A ] = 0 . \label{commutator_gammaTA}
\end{equation}

Among the products of $\mGamma_i$ and $T_A$,
$\mGamma_{ij}$ and $\mGamma_i T_A$ are antisymmetric matrices,
whereas $\mGamma_{ij} T_A$ is symmetric.
In particular, $\mGamma_{ij}$ realizes the spinor representation of
$\mathfrak{so}(5)$ algebra.
Since $\mGamma_i$ and $T_A$ commute,
the $\mathfrak{so}(5)$ and $\mathfrak{su}(2)$ algebras
commute with each other.
The SU(2) group associated with this $\mathfrak{su}(2)$ algebra is
the symmetry with respect to which the symplectic reduction is performed,
whereas the $\mathfrak{so}(5)$ algebra corresponds to
the residual rotational symmetry in five dimensions.

The matrices $\mGamma_{ij}$, $T_A$, $\mGamma_i T_A$ form
a complete basis of antisymmetric $8 \times 8$ matrices.
The number of their independent components is $10 + 3 + 15 = 28$,
which coincides with the dimension of the space of antisymmetric $8 \times 8$
matrices.
These matrices provide a representation of the $\mathfrak{so}(8)$ algebra.
Indeed, defining $M_{ab} \, (a,\,b = 1,\ldots,8)$ by
\begin{align}
  M_{ij} &= - \frac{1}{2} \mGamma_{ij} , \\
  M_{5 + A, 5 + B} &= - \frac{1}{2} \varepsilon_{ABC} T_C , \\
  M_{i, 5 + A} &= - \frac{1}{2} \mGamma_i T_A,
\end{align}
we find that $M_{ab}$ satisfies the $\mathfrak{so}(8)$ algebra
\begin{equation}
  [ M_{ab},\, M_{cd} ] = \delta_{ac} M_{bd} - \delta_{ad} M_{bc}
    - \delta_{bc} M_{ad} + \delta_{bd} M_{ac} .
\end{equation}
Therefore, the eight-dimensional angular momentum $\mu_{ab}^\mathrm{A}$
can be decomposed in terms of $\mGamma_{ij}$, $T_A$ and
$\mGamma_i T_A$ as
\begin{equation}
  \mu_{ab}^\mathrm{A} = \frac{1}{8} \mu_{ij} (\mGamma_{ij})_{ab}
    + \frac{1}{4} \mu_A (T_A)_{ab} + \frac{1}{4} \mu_{i,A} (\mGamma_i T_A)_{ab} ,
\end{equation}
where $\mu_{ij}$, $\mu_A$ and $\mu_{i,A}$ are given by
\begin{align}
  \mu_{ij} &= q^\rT \mGamma_{ij} p , \label{mu_ij} \\
  \mu_A &= q^\rT T_A p , \label{mu_A} \\
  \mu_{i,A} &= q^\rT \mGamma_i T_A p . \label{mu_iA}
\end{align}
In particular, $\mu_{ij}$ is the moment map associated with the SO(5) symmetry
and gives rise to the five-dimensional angular momentum
after the symplectic reduction,
whereas $\mu_A$ is the moment map associated with the SU(2) symmetry
and is the moment map used in the symplectic reduction,
as we shall see in the next section.

On the other hand, the matrices $\mGamma_i$ and $\mGamma_{ij} T_A$
form a complete basis for the space of traceless symmetric $8 \times 8$
matrices.
The number of their independent components is $5+30=35$,
which coincides with the dimension of the space of
traceless symmetric $8 \times 8$ matrices.
The eight-dimensional JHF tensor $\tilde{\mu}_{ab}^\mathrm{S}$
can then be decomposed in terms of $\mGamma_i$ and $\mGamma_{ij} T_A$ as
\begin{equation}
  \tilde{\mu}_{ab}^\mathrm{S} = \frac{1}{8} \mu_i (\mGamma_i)_{ab}
    + \frac{1}{16} \mu_{ij,A} (\mGamma_{ij} T_A)_{ab} ,
\end{equation}
where $\mu_i$ and $\mu_{ij,A}$ are given by
\begin{align}
  \mu_i &= \frac{1}{2 \lambda} (\lambda^2 q^\rT \mGamma_i q +
    p^\rT \mGamma_i p ), \label{mu_i} \\
  \mu_{ij,A} &= \frac{1}{2 \lambda} (\lambda^2 q^\rT \mGamma_{ij} T_A q +
    p^\rT \mGamma_{ij} T_A p ) . \label{mu_ijA}
\end{align}

Together with the identity matrix $\mathbbm{1}_8$,
the matrices \{$\mGamma_{ij}$, $T_A$, $\mGamma_i T_A$\}
and \{$\mGamma_i$, $\mGamma_{ij} T_A$\}
form a complete basis for the space of $8 \times 8$ matrices.

The discussion so far can be summarized in terms of the decomposition of
the SO(8) representations into irreducible representations of
the subgroup $\mathrm{SO}(5)\times \mathrm{SU}(2)$ as
\begin{align}
  \mathbf{28} &=
    (\mathbf{10},\,\mathbf{1}) \oplus (\mathbf{1},\,\mathbf{3})
    \oplus(\mathbf{5},\,\mathbf{3}), \\
  \mathbf{35} &=
    (\mathbf{5},\, \mathbf{1}) \oplus (\mathbf{10},\mathbf{3}) .
\end{align}
With these decompositions, the SU(2)-singlet representations
$(\mathbf{10},\,\mathbf{1})$ and $(\mathbf{5},\,\mathbf{1})$
correspond respectively to $\mGamma_{ij}$ and $\mGamma_i$,
which do not involve the $\mathfrak{su}(2)$ generators $T_A$,
and give rise to SU(2)-invariant conserved quantities
that descend to the reduced phase space.

The actions on the phase space
$\mathbb{R}^8\times\mathbb{R}^8$
corresponding to $\mGamma_{ij}$ and $\mGamma_i$
are given by $\mathbbm{1}_2 \otimes \mGamma_{ij}$ and
$( i \sigma_2) \otimes \mGamma_i$.
Defining $\mathfrak{M}_{IJ}$ $(I,\,J = 1, \ldots, 6)$ by
\begin{equation}
  \mathfrak{M}_{ij} = - \frac{1}{2} \mathbbm{1}_2 \otimes \mGamma_{ij} ,
  \quad
  \mathfrak{M}_{i6} = \frac{1}{2} ( i \sigma_2) \otimes \mGamma_i ,
\end{equation}
we find that $\mathfrak{M}_{IJ}$ satisfies the same $\mathfrak{so}(6)$ algebra
as $L_{IJ}$ given in (\ref{so(6)algebra}).
\begin{equation}
  [ \mathfrak{M}_{IJ} ,\, \mathfrak{M}_{KL} ]
    = \delta_{IK} \mathfrak{M}_{JL} - \delta_{IL} \mathfrak{M}_{JK}
    - \delta_{JK} \mathfrak{M}_{IL} + \delta_{JL} \mathfrak{M}_{IK} .
    \label{so(6)algebra2}
\end{equation}
The $\mathfrak{so}(6)$ algebra generated by $\mathfrak{M}_{IJ}$
also commutes with $\mathfrak{su}(2)$,
extending the commutativity established above for $\mathfrak{so}(5)$.
Consequently, the corresponding SO(6) symmetry,
which is part of the SU(8) dynamical symmetry
of the eight-dimensional harmonic oscillator,
survives the SU(2) symplectic reduction.
As will be shown in section \ref{RL},
the SO(6) moment map gives rise to the angular momentum
and the suitably normalized RL vector of the five-dimensional Kepler problem.
In particular, the normalized RL vector is obtained from the moment map $\mu_i$
in (\ref{mu_i}) associated with $\mathfrak{M}_{i6}$.

It should be emphasized that, in the following calculations,
we do not use the explicit forms of $\mGamma_i$ and $T_A$,
but only relations (\ref{Clifford}), (\ref{su2algebra})
and (\ref{commutator_gammaTA}) together with the Fierz identities,
which follow from the completeness of the matrix basis and are described
in detail in appendix \ref{FierzIdentity}.

\section{Derivation of 5D Kepler problem by symplectic reduction}
\label{derivation}

In this section, we derive the five-dimensional Kepler problem
from the eight-dimensional harmonic oscillator by symplectic reduction.
The mathematical framework of this construction is
provided by the MW theorem~\cite{MW}, which we briefly review below.

The MW theorem is stated as follows.
Let $(M,\,\omega)$ be a symplectic manifold on which a Lie group $G$
act with the moment map $\mu : M \to \mathfrak{g}^\ast$.
Let $i : \mu^{-1}(0) \hookrightarrow M$ denote the inclusion.
Suppose that $G$ acts freely on $\mu^{-1}(0)$. Then,
\begin{enumerate}
  \item The orbital space $M_\mathrm{red} = \mu^{-1}(0)/G$ is a manifold.
  \item The projection $\pi : \mu^{-1}(0) \to M_\mathrm{red}$
    is a principal $G$-bundle.
  \item There exists a symplectic form $\omega_\mathrm{red}$ on $M_\mathrm{red}$
    such that $i^\ast \omega = \pi^\ast \omega_\mathrm{red}$.
\end{enumerate}
$(M_\mathrm{red},\, \omega_\mathrm{red})$ is called the symplectic quotient of
$(M,\,\omega)$.
In other words, the MW theorem realizes a symplectic reduction
of the phase space by fixing the moment map and quotienting
by the corresponding symmetry group.

The reduction from the eight-dimensional harmonic oscillator
to the five-dimensional Kepler problem
is realized by the Hurwitz transformation
\begin{equation}
  x_i = \frac{1}{2} q^\rT \mGamma_i q . \label{Hurwitz_x}
\end{equation}
It is an extension of the Kustaanheimo-Stiefel transformation
to the five-dimensional Kepler problem.
The Hurwitz transformation is invariant under the SU(2) transformation
\begin{equation}
  q \longmapsto e^{- \theta_A T_A} q .
\end{equation}
Hence, the symmetry group $G$ appearing in the MW theorem is
identified with this SU(2), and the corresponding moment map is
$\mu_A$ in (\ref{mu_A}).
Furthermore, by the Fierz identity (\ref{Euleridentity}) yields
\begin{equation}
  x_i^2 = \frac{1}{4} (q^\rT \mGamma_i q)^2
    = \frac{1}{4} (q^\rT q)^2 .
\end{equation}

Let $\tau$ and $t$ denote the time parameters of the
eight-dimensional harmonic oscillator and the five-dimensional Kepler problem,
respectively, and relate them by
\begin{equation}
  dt = (q^\rT q) d \tau . \label{timerescale}
\end{equation}
This form of time reparametrization is standard
in the Kustaanheimo-Stiefel regularization~\cite{KS,SS,Saha}.
Taking the above relation into account,
the Hamilton equation derived from the Hamiltonian (\ref{8dimHOH}),
\begin{equation}
  \frac{d}{d\tau} q_a = \frac{\partial h}{\partial p_a}
    = p_a ,
\end{equation}
can be rewritten as
\begin{equation}
  \frac{d}{dt}{q}_a = \frac{1}{q^\rT q} p_a .
\end{equation}
The corresponding momentum $\pi_i$ of the five-dimensional Kepler problem
is therefore given by
\begin{align}
  \pi_i &= \frac{d}{dt} x_i \notag \\
  &= q^\rT \mGamma_i \frac{d}{dt} q \notag \\
  &= \frac{q^\rT \mGamma_i p}{q^\rT q} . \label{Hurwitz_pi}
\end{align}
Squaring this expression, we obtain
\begin{align}
  \pi_i^2 &= \frac{(q^\rT \mGamma_i p)^2}{(q^\rT q)^2} \notag \\
  &= \frac{(q^\rT q) (p^\rT p) - (q^\rT T_A p)^2}{(q^\rT q)^2} \notag \\
  &= \frac{p^\rT p}{q^\rT q} - \frac{\mu_A^2}{(q^\rT q)^2} , \label{pi^2}
\end{align}
where the second equality follows from the Fierz identity (\ref{8x8Fierz_GT1}).
Using this result, the Hamiltonian of the eight-dimensional harmonic oscillator
(\ref{8dimHOH}) can be rewritten as
\begin{align}
  h &= \frac{1}{2} (q^\rT q) \pi_i^2 + \frac{\mu_A^2}{2 q^\rT q}
    + \frac{1}{2} \lambda^2 q^\rT q \notag \\
  &= \sqrt{x_i^2} \pi_i^2 + \frac{\mu_A^2}{4 \sqrt{x_i^2}}
    + \lambda^2 \sqrt{x_i^2} .  
\end{align}
Restricting to the energy surface $h = 2k$
and setting $\lambda^2 = -2 E$, we obtain
\begin{equation}
  E = - \frac{1}{2} \lambda^2
    = \frac{1}{2} \pi_i^2 - \frac{k}{\sqrt{x_i^2}}
    + \frac{\mu_A^2}{8 x_i^2} . \label{replacelambda^2}
\end{equation}
Here, the identifications $h = 2k$ and $\lambda^2 = -2 E$
reflect the interchange of the roles of the mechanical energy
and the coupling constant associated with the fixed-energy condition
and the time reparametrization (\ref{timerescale}).
With these identifications and the SU(2) moment map constraint $\mu_A = 0$,
the resulting expression
reduces to the five-dimensional Kepler Hamiltonian (\ref{5dimKeplerH}),
with $E$ as the mechanical energy.
Further details of the interchange between the energy and the coupling constant
are given in appendix \ref{interchange}.

We now verify explicitly that the symplectic forms
are compatible with the reduction.
Substituting the Hurwitz transformations (\ref{Hurwitz_x})
and (\ref{Hurwitz_pi}) into the canonical 1-form
$\theta_\mathrm{red} = \pi_i d x_i$ on $\mathbb{R}^5 \times \mathbb{R}^5$,
we obtain
\begin{align}
  \theta_\mathrm{red} &= \frac{q^\rT \mGamma_i p}{q^\rT q}
    d \left( \frac{1}{2} q^\rT \mGamma_i q \right) \notag \\
  &= \frac{1}{2} (\mGamma_i)_{ab} (\mGamma_i)_{cd}
    \frac{q_a p_b}{q^\rT q} d (q_c q_d) .
\end{align}
Using the relation
\begin{equation}
  \delta_{ab} \delta_{cd} + (\mGamma_i)_{ab} (\mGamma_i)_{cd}
    - \frac{1}{2} (\mGamma_{ij})_{ab} (\mGamma_{ij})_{cd}
    = 2 \delta_{ad} \delta_{cb} - 2 (T_A)_{ad} (T_A)_{cb} ,
\end{equation}
which follows from the Fierz identities (\ref{8x8Fierz1}), (\ref{8x8Fierz2})
and (\ref{8x8Fierz3}), this expression can be rewritten as
\begin{align}
  \theta_\mathrm{red} &= p^\rT dq - \frac{1}{q^\rT q}
    (q^\rT T_A p) (q^\rT T_A dq) \notag \\
  &= \theta - \mu_A \frac{q^\rT T_A dq}{q^\rT q} .
\end{align}
Here, $\theta = p_a dq_a = p^\rT dq$ is the canonical 1-form
on  $\mathbb{R}^8 \times \mathbb{R}^8$.
The symplectic form $\omega_\mathrm{red} = dx_i \wedge d \pi_i
= - d\theta_\mathrm{red}$
is therefore given by
\begin{align}
  \omega_\mathrm{red} &= - d \left( \theta - \mu_A \frac{q^\rT T_A dq}{q^\rT q}
    \right) \notag \\
  &= \omega + d \left( \mu_A \frac{q^\rT T_A dq}{q^\rT q} \right) . 
\end{align}
Thus, imposing the SU(2) moment map constraint $\mu_A = 0$,
we obtain the relation $i^* \omega = \pi^* \omega_\mathrm{red}$,
as guaranteed by the MW theorem.

\section{RL vector as SO(6) moment map}
\label{RL}

As shown in section \ref{8x8completebasis},
the generators $\mathfrak{M}_{IJ}$ defined in (\ref{so(6)algebra2}),
constructed from $\mathbbm{1}_2 \otimes \mGamma_{ij}$
and $(i \sigma_2) \otimes \mGamma_i$,
satisfy the $\mathfrak{so}(6)$ algebra.
We now derive the normalized five-dimensional RL vector from the corresponding
SO(6) moment map.
We first derive the five-dimensional angular momentum $L_{ij}$
from the SO(5) moment map $\mu_{ij}$ given in (\ref{mu_ij}).
Inserting the identity matrix given
in the form (\ref{identity2}) into $\mu_{ij}$, we obtain
\begin{align}
  \mu_{ij} &= q^\rT \mGamma_{ij} \mathbbm{1}_8 p \notag \\
  &= q^\rT \mGamma_{ij} \frac{\mGamma_k (qq^\rT) \mGamma_k
    - T_A (qq^\rT) T_A}{q^\rT q} p \notag \\
  &= (q^\rT \mGamma_i q)\frac{q^\rT \mGamma_j p}{q^\rT q}
    - (q^\rT \mGamma_j q)\frac{q^\rT \mGamma_i p}{q^\rT q}
    - (q^\rT T_A p) \frac{q^\rT \mGamma_{ij} T_A q}{q^\rT q} \notag \\
  &= 2 x_i \pi_j - 2 x_j \pi_i - \mu_A \frac{q^\rT \mGamma_{ij} T_A q}{q^\rT q}
    \notag \\
  &= 2 L_{ij} - \mu_A \frac{q^\rT \mGamma_{ij} T_A q}{q^\rT q} , \label{muL}
\end{align}
which give rise to the five-dimensional angular momentum $L_{ij}$ upon imposing
the SU(2) moment map constraint $\mu_A = 0$.

Next, we derive the normalized RL vector $\tilde{A}_i$
from the moment map $\mu_i$ in (\ref{mu_i})
associated with $\mathfrak{M}_{i6}$.
Eliminating $\lambda^2$ using (\ref{replacelambda^2}),
the first term of $\lambda \mu_i$ becomes
\begin{equation}
  \frac{1}{2} \lambda^2 q^\rT \mGamma_i q
    = \left( - \pi_k^2 + \frac{2k}{\sqrt{x_k^2}} - \frac{\mu_A^2}{4 x_k^2}
    \right) x_i .
\end{equation}
The second term of $\lambda \mu_i$ can be evaluated by
inserting the identity matrix in the form given in (\ref{identity1})
and simplifying the resulting expression.
\begin{align}
  \frac{1}{2} p^\rT \mGamma_i p &=
    \frac{1}{2} p^\rT \mathbbm{1}_8 \mGamma_i p \notag \\
  &= \frac{1}{2} p^\rT \frac{- 2 qq^\rT + 2 \mGamma_k (qq^\rT) \mGamma_k
    + (q^\rT \mGamma_k q) \mGamma_k}{q^\rT q} \mGamma_i p \notag \\
  &= \frac{q^\rT \mGamma_k p}{q^\rT q} (q^\rT \mGamma_{ki} p)
    + \frac{1}{2} (q^\rT \mGamma_i q) \frac{p^\rT p}{q^\rT q} \notag \\
  &= - \pi_k \left( 2 L_{ik} - \mu_A \frac{q^\rT \mGamma_{ik} T_A q}{q^\rT q}
    \right) + x_i \left( \pi_k^2 + \frac{\mu_A^2}{4 x_k^2} \right) ,
\end{align}
where the last equality follows from (\ref{Hurwitz_pi}), (\ref{pi^2})
and (\ref{muL}).
Combining these results, we obtain
\begin{align}
  \lambda \mu_i =& \, \left( -\pi_k^2 + \frac{2k}{\sqrt{x_k^2}}
    - \frac{\mu_A^2}{4 x_k^2} \right) x_i \notag \\
  & - \pi_k \left( 2 L_{ik} - \mu_A \frac{q^\rT \mGamma_{ik} T_A q}{q^\rT q}
    \right) + x_i \left( \pi_k^2 + \frac{\mu_A^2}{4 x_k^2} \right) \notag \\
  =& \, - 2 \left( \pi_k L_{ik} - \frac{k}{\sqrt{x_k^2}} x_i \right)
    + \mu_A \pi_k \frac{q^\rT \mGamma_{ik} T_A q}{q^\rT q} \notag \\
  =& \, - 2 A_i + \mu_A \pi_k \frac{q^\rT \mGamma_{ik} T_A q}{q^\rT q} ,
\end{align}
which give rise to the five-dimensional RL vector $A_i$
upon imposing $\mu_A = 0$.
For $E < 0$, the identification $\lambda^2 = -2 E$, together with $\lambda> 0 $,
gives $\lambda = \sqrt{- 2 E}$. Thus,
\begin{align}
  \mu_i &= - 2 \frac{A_i}{\sqrt{- 2 E}}
    + \frac{\mu_A \pi_k}{\sqrt{- 2 E}} \frac{q^\rT \mGamma_{ik} T_A q}{q^\rT q}
    \notag \\
  &= - 2 \tilde{A}_i + \frac{\mu_A \pi_k}{\sqrt{- 2 E}}
    \frac{q^\rT \mGamma_{ik} T_A q}{q^\rT q} .
\end{align}
Therefore, the normalized five-dimensional RL vector $\tilde{A}_i$ is obtained
from the SO(6) moment map under the condition $\mu_A = 0$
on a fixed negative-energy surface.
Since the SO(6) symmetry underlying this construction is
a subgroup of the SU(8) dynamical symmetry,
this result shows that the normalized five-dimensional RL vector
originates from the SU(8) dynamical symmetry associated
with the geometry of the phase space $\mathbb{R}^8 \times \mathbb{R}^8$.

\section{Summary}
\label{summary}

In this paper, we have investigated the relation
between the conserved quantities of the eight-dimensional harmonic oscillator
and those of the five-dimensional Kepler problem
in terms of classical moment maps and symplectic reduction.
Previous studies have established the relation between the SU(8) symmetry
of the harmonic oscillator and the SU(4) symmetry of the SU(2) Kepler problem,
including the construction of angular momentum
and Runge-Lenz-like operators~\cite{IS1,IS2}.
In the present work, using a complete basis of real $8 \times 8$ matrices
constructed from five-dimensional gamma matrices
and $\mathfrak{su}(2)$ generators,
we have decomposed the conserved quantities of the harmonic oscillator
into representations of $\mathrm{SO}(5) \times \mathrm{SU}(2)$.
The SU(2)-singlet sector generates an SO(6) subgroup
of the SU(8) dynamical symmetry.
We have then shown explicitly that the corresponding SO(6) moment map
gives rise to the angular momentum and the suitably normalized RL vector
of the five-dimensional Kepler problem through symplectic reduction.

Although the analysis presented here has been restricted to
negative mechanical energies $E < 0$,
our method can also be applied to positive energies $E > 0$
by considering an inverted harmonic oscillator with
$\lambda^2 = - \kappa^2 = - 2 E$, where $\kappa > 0$.
In this case, the dynamical symmetry algebra of the eight-dimensional system is
$\mathfrak{sl}(8,\,\mathbb R)$, rather than $\mathfrak{su}(8)$,
after excluding the generator corresponding to the Hamiltonian.
Like $\mathfrak{su}(8)$, this algebra can be constructed from
28 antisymmetric and 35 traceless symmetric $8 \times 8$ real matrices.
Therefore, the decomposition based on gamma matrices
and $\mathfrak{su}(2)$ generators, as well as the Fierz identities
used in the present work, can be applied without modification.
The corresponding SU(2)-invariant generators
form the $\mathfrak{so}(5,\,1)$ algebra associated with
the dynamical symmetry of the five-dimensional Kepler problem
at positive energies.

The RL vector exists in the Kepler problem in arbitrary spatial dimensions,
whereas transformations relating harmonic oscillators to Kepler systems
are known only in two, three, five, and nine dimensions.
For the nine-dimensional case, a correspondence
with the sixteen-dimensional harmonic oscillator
has been proposed~\cite{LeK}.
The transformation can be expressed using nine-dimensional gamma matrices
represented by $16 \times 16$ real symmetric matrices.
Unlike the five-dimensional case, however,
these gamma matrices and their antisymmetrized products
already form a complete basis for the space of $16 \times 16$ real matrices.
Consequently, there are no nontrivial antisymmetric matrices
commuting with all nine gamma matrices
that could generate a continuous symmetry analogous to the SU(2) symmetry
used in the present reduction.
This obstruction is consistent with the octonionic Hopf fibration
$S^{15} \to S^8$, whose fibers are seven-dimensional spheres
rather than Lie groups~\cite{MNY}.

Nevertheless, the algebraic structure
of the sixteen-dimensional harmonic oscillator
contains an $\mathfrak{so}(10)$ subalgebra analogous to
the $\mathfrak{so}(6)$ subalgebra identified in the present work.
In particular, the 36 antisymmetrized products $\mGamma_{ij}$
and the nine gamma matrices $\mGamma_i$ can be used to construct
generators of $\mathfrak{so}(10)$ through their appropriate actions
on the phase space of the sixteen-dimensional harmonic oscillator.
Thus, the difficulty in extending our method
to the nine-dimensional Kepler problem lies
not in the absence of the relevant dynamical symmetry algebra,
but in the absence of a corresponding symmetry subgroup of SO(16)
that would allow the corresponding symplectic reduction.
This suggests that the relation between the $\mathfrak{so}(10)$ moment map
and the nine-dimensional RL vector may
require a different approach to reduction.

\appendix

\section{Fierz identities}
\label{FierzIdentity}

As explained in section \ref{8x8completebasis},
a complete basis for the space of $8\times 8$ real matrices
is given by the five-dimensional gamma matrices $\mGamma_i$,
the generators $T_A$ of $\mathfrak{su}(2)$, and their products.
For convenience, we collectively denote these 64 basis matrices by
$\{X_\alpha\}$ ($\alpha=1,\ldots, 64$).
They satisfy the orthogonality relation
\begin{equation}
  \operatorname{tr} (X_\alpha^\rT X_\beta) = 8 \delta_{\alpha\beta} .
\end{equation}
For general $8\times 8$ real matrices $A$ and $B$,
regarding $A_{ab} B_{cd}$ as a matrix in the indices $c$, $b$
for fixed $a$ and $d$, we can expand it in this basis as
\begin{equation}
  A_{ab} B_{cd} = \sum_{\alpha = 1}^{64} (C_\alpha)_{ad} (X_\alpha)_{cb} ,
    \label{8x8expansion}
\end{equation}
where $C_\alpha$ are $8\times 8$ real matrices.
The coefficients $C_\alpha$ can be determined
using the orthogonality relation above.
Applying this expansion (\ref{8x8expansion}) to the basis matrices,
we obtain the following Fierz identities.
\begin{align}
    \delta_{ab} \delta_{cd} =& \, \frac{1}{8} \left (\delta_{ad} \delta_{cb}
    - (T_A)_{ad} (T_A)_{cb} \right) \notag \\
  & + \frac{1}{8} \left( (\mGamma_i)_{ad}  (\mGamma_i)_{cb}
      - (\mGamma_i T_A)_{ad}  (\mGamma_i T_A)_{cb} \right) \notag \\
  & - \, \frac{1}{16} \left( (\mGamma_{ij})_{ad} (\mGamma_{ij})_{cb}
    - (\mGamma_{ij} T_A)_{ad} (\mGamma_{ij} T_A)_{cb} \right) ,  \label{8x8Fierz1} \\
  (\mGamma_i)_{ab} (\mGamma_i)_{cd} =& \,
    \frac{5}{8} \left( \delta_{ad} \delta_{cb}
    - (T_A)_{ad} (T_A)_{cb} \right) \notag \\
  & - \frac{3}{8} \left( (\mGamma_i)_{ad} (\mGamma_i)_{cb}
      - (\mGamma_i T_A)_{ad} (\mGamma_i T_A)_{cb} \right) \notag \\
  & - \frac{1}{16} \left(
    (\mGamma_{ij})_{ad} (\mGamma_{ij})_{cb}
      - (\mGamma_{ij} T_A)_{ad} (\mGamma_{ij} T_A)_{cb} \right) , \label{8x8Fierz2} \\
  (\mGamma_{ij})_{ab} (\mGamma_{ij})_{cd} =&\,
    - \frac{5}{2} \left( \delta_{ad} \delta_{cb}
     - (T_A)_{ad} (T_A)_{cb} \right)  \notag \\
  & - \frac{1}{2}
      \left( (\mGamma_i)_{ad} (\mGamma_i)_{cb}
      - (\mGamma_i T_A)_{ad} (\mGamma_i T_A)_{cb} \right) \notag \\
  & - \frac{1}{4} \left(
    (\mGamma_{ij})_{ad} (\mGamma_{ij})_{cb}
      - (\mGamma_{ij} T_A)_{ad} (\mGamma_{ij} T_A)_{cb} \right) ,  \label{8x8Fierz3} \\
  (T_A)_{ab} (T_A)_{cd} =& \, \frac{1}{8} \left (-3 \delta_{ad} \delta_{cb}
    - (T_A)_{ad} (T_A)_{cb} \right) \notag \\
  & + \frac{1}{8} \left( -3 (\mGamma_i)_{ad}  (\mGamma_i)_{cb}
      - (\mGamma_i T_A)_{ad}  (\mGamma_i T_A)_{cb} \right) \notag \\
  & - \frac{1}{16} \left( -3 (\mGamma_{ij})_{ad} (\mGamma_{ij})_{cb}
    - (\mGamma_{ij} T_A)_{ad} (\mGamma_{ij} T_A)_{cb} \right) , \label{8x8Fierz4} \\
  (\mGamma_i T_A)_{ab} (\mGamma_i T_A)_{cd} =& \,
    \frac{5}{8} \left( - 3 \delta_{ad} \delta_{cb}
    - (T_A)_{ad} (T_A)_{cb} \right)  \notag \\
  & - \frac{3}{8} \left( - 3(\mGamma_i)_{ad} (\mGamma_i)_{cb}
    - (\mGamma_i T_A)_{ad} (\mGamma_i T_A)_{cb} \right) \notag \\
  & - \frac{1}{16} \left(
    - 3( \mGamma_{ij})_{ad} (\mGamma_{ij})_{cb}
    - (\mGamma_{ij} T_A)_{ad} (\mGamma_{ij} T_A)_{cb} \right) , \label{8x8Fierz5} \\
  (\mGamma_{ij} T_A)_{ab} (\mGamma_{ij} T_A)_{cd} =& \,
    - \frac{5}{2} \left( -3 \delta_{ad} \delta_{cb}
    -  (T_A)_{ad} (T_A)_{cb} \right) \notag \\
  & - \frac{1}{2}
      \left( - 3 (\mGamma_i)_{ad} (\mGamma_i)_{cb}
      - (\mGamma_i T_A)_{ad} (\mGamma_i T_A)_{cb} \right) \notag \\
  & - \frac{1}{4} \left(
    - 3 (\mGamma_{ij})_{ad} (\mGamma_{ij})_{cb}
      - (\mGamma_{ij} T_A)_{ad} (\mGamma_{ij} T_A)_{cb} \right) . \label{8x8Fierz6}
\end{align}
Several useful relations follow from these identities.
(\ref{8x8Fierz1}) and (\ref{8x8Fierz2}) give
\begin{align}
    \delta_{ab} \delta_{cd} - (\mGamma_i)_{ab} (\mGamma_i)_{cd}
    =& \, - \frac{1}{2} \delta_{ad} \delta_{cb}
    + \frac{1}{2} (\mGamma_i)_{ad} (\mGamma_i)_{cb} \nonumber \\
  & + \frac{1}{2} (T_A)_{ad} (T_A)_{cb} 
    - \frac{1}{2} (\mGamma_i T_A)_{ad} (\mGamma_i T_A)_{cb} . \label{8x8Fierz_dG}
\end{align}
Contracting this relation with $q_a q_d$ and $q_a  q_b q_c q_d$, respectively,
we obtain
\begin{equation}
  q_b q_c - (q^\rT \mGamma_i)_b (\mGamma_i q)_c
    = - \frac{1}{2} (q^\rT q) \delta_{cb}
    + \frac{1}{2} (q^\rT \mGamma_i q) (\mGamma_i)_{cb} ,
    \label{8x8Fierz_dG2}
\end{equation}
and \begin{equation}
  (q^\rT q)^2 - (q^\rT \mGamma_i q)^2 =
    - \frac{1}{2} (q^\rT q)^2 + \frac{1}{2} (q^\rT \mGamma_i q)^2 .
    \label{8x8Fierz_dG1}
\end{equation}
(\ref{8x8Fierz_dG1}) leads to the identity
\begin{equation}
  (q^\rT \mGamma_i q)^2
    = (q^\rT q)^2 , \label{Euleridentity}
\end{equation}
and (\ref{8x8Fierz_dG2}) gives the representation of the
identity matrix
\begin{equation}
  \mathbbm{1}_8 = \frac{- 2 qq^\rT + 2 \mGamma_i (qq^\rT) \mGamma_i
    + (q^\rT \mGamma_i q) \mGamma_i}{ q^\rT q} .
    \label{identity1}
\end{equation}
Similarly, the relation obtained
from (\ref{8x8Fierz2}) and (\ref{8x8Fierz4}) is
\begin{align}
  &(\mGamma_i)_{ab} (\mGamma_i)_{cd}
    - (T_A)_{ab} (T_A)_{cd}  \notag \\ 
  &= \delta_{ad} \delta_{cb}
    - \frac{1}{2}(T_A)_{ad} (T_A)_{cb} 
    + \frac{1}{2} (\mGamma_i T_A)_{ad} (\mGamma_i T_A)_{cb}
   - \frac{1}{4} 
  (\mGamma_{ij})_{ad} (\mGamma_{ij})_{cb} .
  \label{8x8Fierz_GT}
\end{align}
Contracting this relation with $q_a q_d$ and $q_a p_b p_c q_d$, respectively,
we obtain,
\begin{equation}
  (q^\rT \mGamma_i)_b (\mGamma_i q)_c
    - (q^\rT T_A)_b (T_A q)_c = (q^\rT q) \delta_{cb} , \label{8x8Fierz_GT2}
\end{equation}
and
\begin{equation}
  (q^\rT \mGamma_i p)^2 + (q^\rT T_A p)^2 =
    (q^\rT q) (p^\rT p) . \label{8x8Fierz_GT1}
\end{equation}
From (\ref{8x8Fierz_GT2}),
we obtain another representation of the identity matrix
\begin{equation}
  \mathbbm{1}_8 = \frac{ \mGamma_i (qq^\rT)
    \mGamma_i - T_A  (qq^\rT) T_A}{q^\rT q} . \label{identity2}
\end{equation}

\section{Interchange between energy and coupling in the Hurwitz transformation}
\label{interchange}

In this appendix, we examine in more detail the interchange
between the energy and the coupling constant
represented by (\ref{replacelambda^2})
in the transformation from the harmonic oscillator to the Kepler problem.
In particular, we clarify its relation to the time reparametrization
by comparing the corresponding Hamiltonian vector fields.

For the Hamiltonian $h$ of the harmonic oscillator,
we introduce a new Hamiltonian $K$ by
\begin{equation}
  K = \frac{1}{q^\rT q} ( h - \varepsilon ) , \label{K}
\end{equation}
where $\varepsilon$ is a constant specifying the energy surface
$h = \varepsilon$ considered below.
The factor $1/(q^\rT q)$ in (\ref{K}) corresponds to
the time reparametrization (\ref{timerescale}),
which relates the time parameter $\tau$ of the harmonic oscillator
to the time parameter $t$ associated with $K$.
Indeed, integrating $K$ with respect to $t$, we obtain
\begin{equation}
  \int K dt = \int \frac{1}{q^\rT q} ( h - \varepsilon ) dt
  = \int ( h - \varepsilon ) d\tau .
\end{equation}
Thus, the factor $1/(q^\rT q)$ in $K$ accounts for the time reparametrization.
To compare the trajectories generated by $K$ and $h$,
we now consider the relation between their Hamiltonian vector fields.
The Hamiltonian vector field $X_K$ associated with $K$
is related to the Hamiltonian vector field $X_h$ associated with $h$ by
\begin{equation}
  X_K = \frac{1}{q^\rT q} X_h
    + (h - \varepsilon) \frac{2 q_a}{(q^\rT q)^2} \frac{\partial}{\partial p_a}.
\end{equation}
On the energy surface $h = \varepsilon$, this relation reduces to
\begin{equation}
  X_K = \frac{1}{q^\rT q} X_h . \label{XKXh}
\end{equation}
Thus, on this energy surface,
the integral curves of the two Hamiltonian vector fields
coincide up to a reparametrization.
Consequently, $K$ and $h$ generate the same trajectories.

Substituting the explicit expression (\ref{8dimHOH}) for $h$
and the Hurwitz transformations (\ref{Hurwitz_x}) and (\ref{Hurwitz_pi})
into $K$, we obtain
\begin{align}
  K &= \frac{1}{2} \frac{p^\rT p}{q^\rT q} + \frac{1}{2} \lambda^2
    - \frac{\varepsilon}{q^\rT q} \notag \\
  &= \left( \frac{1}{2} \pi_i^2 + \frac{\mu_A^2}{8 x_i^2}
    - \frac{\varepsilon}{2 \sqrt{x_i^2}} \right)
    + \frac{1}{2} \lambda^2 \notag \\
  &= H_\mu + \frac{1}{2} \lambda^2 ,
\end{align}
where $H_\mu$ is defined by
\begin{equation}
  H_\mu = \frac{1}{2} \pi_i^2 + \frac{\mu_A^2}{8 x_i^2}
    - \frac{\varepsilon}{2 \sqrt{x_i^2}} .
\end{equation}
This relation shows that the energy surface $h = \varepsilon$
corresponds to the energy surface $H_\mu = - \frac{1}{2} \lambda^2$.
Setting $\varepsilon = 2k$ and
imposing the SU(2) moment map constraint $\mu_A=0$,
$H_\mu$ reduces to the Hamiltonian $H$ of the Kepler problem
(\ref{5dimKeplerH}).
As shown in (\ref{XKXh}), the Hamiltonian vector fields $X_K$ and $X_h$
differ by the factor $1/(q^\rT q)$ on the energy surface $h = 2k$.
Since the preservation of the symplectic form under the reduction
is guaranteed by the MW theorem, the corresponding relation holds
for $X_H$ under the condition $\mu_A = 0$.
\begin{equation}
  X_H = \frac{1}{2 \sqrt{x_i^2}} X_h .
\end{equation}
Here, the right-hand side is understood as the projection
of the vector field onto the reduced phase space.
Thus, up to the time reparametrization,
the trajectories of the harmonic oscillator
are mapped to those of the Kepler problem under the reduction,
with the energy $\varepsilon$ of the harmonic oscillator
becoming the coupling constant $k = \frac{1}{2} \varepsilon$
of the Kepler problem,
while the coupling constant $\lambda^2$ of the harmonic oscillator
determines the energy $E = - \frac{1}{2} \lambda^2$ of the Kepler problem.

This interchange can be regarded as an instance of
coupling-constant metamorphosis,
in which the roles of a coupling constant and the energy are
exchanged through a transformation
involving a reparametrization of time~\cite{HGDR}.
Coupling-constant metamorphosis is
a special case of the St\"ackel transform.
The St\"ackel transform maps one integrable Hamiltonian system
to another while preserving integrability~\cite{BKM}.
In this framework, the accompanying time reparametrization can be
viewed as the corresponding reciprocal transformation~\cite{SB}.

\end{document}